# Neural network facilitated ab initio derivation of linear formula: A case study on formulating the relationship between DNA motifs and gene expression

Chengyu Liu[1], Wei Wang[1,2*]

[1]Department of Chemistry and Biochemistry, [2]Department of Cellular and Molecular Medicine, University of California, San Diego, La Jolla, CA, USA.

*Corresponding author: Wei Wang, E-mail: wei-wang@ucsd.edu

## Abstract

Developing models with high interpretability and even deriving formulas to quantify relationships between biological data is an emerging need. We propose here a framework for ab initio derivation of sequence motifs and linear formula using a new approach based on the interpretable neural network model called contextual regression model. We showed that this linear model could predict gene expression levels using promoter sequences with a performance comparable to deep neural network models. We uncovered a list of 300 motifs with important regulatory roles on gene expression and showed that they also had significant contributions to cell-type specific gene expression in 154 diverse cell types. This work illustrates the possibility of deriving formulas to represent biology laws that may not be easily elucidated. (https://github.com/Wang-lab-UCSD/Motif_Finding_Contextual_Regression)

## Introduction

Deep neural networks have been widely adopted in modeling biological data and they can achieve state-of-the-art prediction accuracy in different datasets such as gene expression, open chromatin and histone marks[1–7]. For example, convolutional neural network (CNN)[8] is suitable for processing sequence data as the filters in the CNN, which CNN uses to find sequence patterns for prediction, can represent DNA motifs and other functional patterns. A limitation of the highly complex and non-linear nature of neural network models is their low interpretability[9]. For instance, as the data go through multiple layers of matrix multiplication and non-linear activation functions in CNN, the contribution of each input feature to prediction result is hardly tractable and the filters also unlikely have equal contribution to the predictions. To address this challenge, methods such as LIME[10], SHAP[11], Contextual Regression (CR)[6] and DeepLift[12] have been developed to identify features important for the prediction performance. Such analyses are powerful and they can provide informative insights, such as which base-pairs are important in a sequence prediction task[4].

Despite these progresses, a tempting goal not yet achieved for building interpretable models is to derive formulas, which is intuitive for interpretation and human comprehension, for analytical quantification of the relationship between biological data such as DNA sequence and gene expression. A classic approach to derive formulas from data is symbolic regression[13] that searches combinations of predefined operations and functions, such as addition, subtraction, multiplication, division and exponent, on input variables to predict output variables. For example, symbolic regression[13] and, more recently, symbolic regression combined with neural networks[14] have been shown to successfully recover physics laws from data. These studies are inspiring to investigate the possibility of deriving biology laws. However, it is imaginable that living systems are more complex than the physical subjects and in many cases defining appropriate input variables predictive of a target observation is a great challenge; in fact the complexity of the DNA sequence and the exponentially increasing searching space of operation combinations have hindered broad application of similar approaches in interpreting biological data.

In this study, we propose a framework to derive the simplest formulas, linear formulas, to model biological relationships. As many physics laws, such as Newton's second law F=ma and mass-energy equivalence E=mC$^2$, are in linear formulas, it is reasonable to argue some biology laws may also be linear[15] or can be well approximated by linear relationship[16–18]. We conducted a proof-of-concept study on predicting gene expression using promoter sequences. Promoters are crucial for regulating gene transcription[19]. DNA motifs, particularly those in the core promoter regions[20], that are important for transcriptional regulation have been investigated from molecular, genetic and functional aspects[21–36].

Many studies have shown that promoter sequences, genomic or synthetic ones, are predictive of gene expression and linear regression models of using sequence features in the promoters have reasonable performance while deep neural networks can achieve better predictions[5]. Such a linear relationship allows testing of our approach on (1) ab initio derivation of the input sequence features (i.e. DNA motifs) and (2) building a linear model on these derived features to achieve prediction performance comparable to deep neural networks.

**Results**

**Ab initio derivation of linear model of sequence motifs to predict gene expression levels.**

A roadblock towards deriving a linear model to predict gene expression from promoter sequences is to uncover appropriate input features. DNA motifs represent genomic segments recognized by TFs and specific TFs binding to a particular promoter is the key of regulating transcription. Therefore, a linear model of DNA motifs to predict gene expression would naturally quantify the contribution of motifs and reveal the important transcriptional regulators. Our approach started with using a CNN model based on the contextual regression framework[6] that can learn expression related motifs and predict expression level by generating local linear models. The motifs and their occurrence counts were directly learned by this contextual regression model, and then the motif occurrence counts were used to fit a global linear model to predict RNA abundance achieving a performance comparable to pure neural network models.

The contextual regression model aims to simultaneously optimize prediction accuracy and assess the contribution of each input feature to the prediction. It is based on the strength of deep neural networks on end-to-end learning[37] process with automatic feature engineering (Figure 1A). The model consists of two modules: motif count generating module and local linear model generating module. The architecture is demonstrated as shown below using motifs of 3-mers for illustrative purpose (in actuality, we used motifs of 8-mers in the prediction): The sequence is first input into a layer of CNN filters, each filter represents a motif of n-mers (8-mers were used in our model). The activation of each filter in a genomic location represents that the corresponding motif is found there. Then the output is fed into two branches: the first branch applies a sigmoid function (to binarize the output) and sum across the sequence to generate the count of each motif, which we call motif finding branch. The second branch is input into a convolutional neural network that generates the contextual weights (local linear model), which we call contextual weight branch. The outputs from the two branches are then dot-producted (equivalent to applying the local linear model) to yield the prediction of expression level.

This way the CNN was trained on the promoter sequences and gene expression data to find local linear models for individual data points so that the two fitting problems were solved simultaneously: accurate prediction of the output (i.e. gene expression) and assessing the contribution of individual features represented by the contextual weights. We downloaded the gene expression data, which is the median expression level of 56 cell types, from ref. [5]. To assess the model performance, we randomly divided the data 10 times to generate 10 datasets for accuracy evaluation.

Human promoters include sequences upstream and downstream of transcription start site (TSS). Different lengths of sequences have been used in the literature, ranging widely from (-1000bp, +500bp) to (-7000bp, +3000bp). We thus first aimed to identify a reasonably short length of sequences around TSS that can predict gene expression with satisfactory performance, because relative short sequences would reduce the possibility of overfitting and alternative models, making it more straightforward to compare the contextual regression model with the pure neural network models. We assessed the prediction accuracy of our model on three different sizes around TSS: (-100bp, +100bp), (-200bp, +200bp) and (-1000bp, 1000bp). We found the performance (Pearson correlation R between predicted and experimental values) of (-200bp, +200bp) (0.675±0.018) was very similar to (-1000bp, 1000bp) (0.682±0.024) and yet sufficiently higher than (-100bp, +100bp) (0.648±0.019) (Figure 1.B). For a comparison, a slightly higher Pearson correlation could be achieved on a much longer region of (-7000bp, +3500bp) around TSS either using a deep neural network model in ref. [5] (Pearson R=0.71) or a contextual regression model with the same model structure using (-1000bp, 1000bp) (Pearson R=0.68, Figure 1A). It is obvious that the longer the promoter region, the more information obtained for more accurate prediction of gene expression. For our purpose of demonstrating the feasibility of deriving linear formulas, we chose (-200bp, +200bp) around TSS to train the model and the follow up analyses in the rest of the paper.

Then we performed the second step to generate the global linear model. A global linear model was fit for each of the 10 runs on the 10 random partitions of the data (Figure 1.C). The global linear models show slightly worse but still highly compatible performance (R=0.65±0.02) compared to the contextual regression model (R=0.68±0.02) and the Xpresso model structure (R=0.67±0.02), suggesting that the RNA expression level can be predicted using linear models of sequence motifs (Figure 1D).

To investigate whether the contextual regression model helped to uncover appropriate features, we developed a benchmark neural network model which predicts gene expression solely using motif counts (called motif finding NN): this model is the motif finding branch of the contextual regression model connected to a 3-layer feed forward neural network model (Figure 1SB). This model achieved a Pearson R=0.52±0.02 which is much worse than contextual regression. This is probably due to the fact that the contextual weight branch of the contextual regression model takes in information from the whole sequence, which facilitates learning. We also built a global linear model to predict gene expression based on the activation count of the filters (i.e. equivalent to the motif counts) in the Xpresso model (i.e. the output of the first layer activated by ReLU summed over each (-200bp, +200bp) region) and the average Pearson R was 0.41±0.03 on the test set. This result indicated that simply replacing neural network with linear regression does not necessarily work and the contextual regression framework facilitates deriving input features that are able to fit a linear model for prediction.

**Uncovering a list of promoter motifs important for regulating gene expression.**

DNA motifs in the human promoters have been well analyzed but new motifs and new insights keep emerging from recent studies[27,29,30,38]. Leveraging the power of the highly interpretable linear model, we set to define a list of promoter motifs that are most important for regulating gene expression. A technical hurdle to overcome is that there exist alternative models to achieve the same prediction accuracy. A common strategy is to combine those models, which can not only unify the models but also increase model adaptivity to different datasets. Thus we created a combined model from the models we generated in the 10 runs. Furthermore, some de novo motifs may match with the known motifs. To reduce the effort of distinguishing known and de novo motifs in the follow up analysis, we chose to include the known motifs in the core vertebrate JASPAR database (a total of 831 motifs)[39] in addition to the 640 de novo motifs found by the 10 contextual regression (CR) models (Figure 2.A). Then we selected the top motifs based on their predictive power on gene expression. We chose the regression tree[40] as our selection model that uses entropy to represent feature significance. We recursively remove 100 motifs in each round until the number of motifs reach our target value.

    We compared the prediction accuracy of linear models using the top selected motifs with a baseline linear regression model directly trained on the JASPAR motifs and the neural network models to predict gene expression. We used the default division of train, validation and test set from ref.[5] for a fair comparison. Among all the models, the one using the top 300 motifs selected from JASPAR+CR Found motifs achieved the best performance on the test set while being linearly interpretable and simpler (Pearson R=0.70), same as the two neural network models and much better than the baseline linear model (R=0.60). On the test set, the top 300 motifs showed better performance than using all the motifs which is likely due to removing low information motifs that can fit to noise on the training set. Note that the top-300-motif model had the same performance on training, testing and whole dataset fitting, while the top-200-motif model, despite achieving close accuracy on the test set, performed slightly worse on training and whole dataset, suggesting some underfitting. Taken together, we chose the linear top-300-motif model for the follow up analysis.

    We noticed that the most high-weight motifs (194 out of 300, mean absolute weight 1.53) came from contextual regression (in fact all top 20 positive and negative weighted motifs were from contextual regression) rather than from the JASPAR database (106 out of 300, mean absolute weight 0.012),

suggesting that the de novo CR motifs are stronger features for expression prediction than the known motifs. This is not surprising since the motifs from contextual regression are directly trained to predict gene expression.

We next examined the spatial distribution of the motifs with the top 20 positive and negative weights in the promoters. The distribution is calculated by averaging the output from the motif_exists module (Figure 1SA) of our model over all sequences in the training, validation and testing set. These motifs exhibit a strong change of location preference near the TSS: for instance, motif R1M4 has an increasing presence when approaching TSS. Near TSS it has a strong preference for specific positions (large fluctuation of distribution). Another example is motif R2M6, which has an even distribution before and after TSS and has a strong preference for specific positions near TSS (Figure 2.B). This suggests certain binding spots near TSS are preferred.

The de novo motifs discovered by contextual regression model are highly effective on predicting transcription levels ande most of them do not match with the known motifs in the JASPAR[39] or CIS-BP[41] database. To infer their cellular function, for each of the top 20 positive and top 20 negative weight motifs, we extracted the top 500 genes with the highest total motif count and performed GO term enrichment analysis[42–44] on their biological process and molecular function. We found three main themes of GO terms associated with these motifs (Figure 2.B): (1) development (for instance, nervous system development (GO:0007399), multicellular organism development (GO:0007275) and system development (GO:0048731)), (2) signaling and regulation (for instance, regulation of signaling (GO:0023051), regulation of cellular process (GO:0050794) and regulation of signal transduction (GO:0009966)) and (3) stimulus-response (for instance, detection of chemical stimulus (GO:0009593), detection of chemical stimulus involved in sensory perception (GO:0050907) and G protein-coupled receptor signaling pathway (GO:0007186)). The association of the high weighted motifs with these housekeeping functions indicated their importance of regulating expression levels of the related genes.

**Predicting gene expression in diverse cell types.**

We next examined whether the top 300 motifs define a parts list of TF binding motifs in the promoters and whether the linear relationship remains valid on predicting cell-type specific gene expressions. We collected the RNA-seq data of 154 cell types/tissues[45] from Epigenomic Roadmap Project[46] and ENCODE[47]. We trained and tested the model using the same division of train, validate and test set as in the previous section. Our model achieved an average Pearson R of 0.61 on the test set for all cell types/tissues. Among the 154 cell lines, 98% (151) of them had a Pearson R above 0.5. Considering that no epigenetic state or enhancer regulation was included in the model, such a performance indicates that the 300 motifs in the promoters play a major role of regulating cell-type specific gene expression. The 3 cell lines that had Pearson R below 0.5 are all adult cell lines (R.adt.sigmoid.colon, R.adt.small.intestine, E.adt.right.lobe.of.liver). Indeed, the average Pearson correlation for the adult cell types/tissues was 0.59, which is significantly lower than 0.61 (p value=0.022) for children and 0.64 (p value=5.6e-7) for the embryonic cells/tissues (Table 1). As gene expressions of the cell types/tissues in the later stages of development are generally more affected by epigenetic modifications and enhancers, this observation highlights the importance to consider information other than promoter sequences in the prediction model.

We next investigated how the top 20 positive and negative weighted motifs selected from predicting the median gene expression values in the previous section are functioning in regulating cell-type specific expressions. We performed hierarchical clustering on the cells/tissues based on their linear model weights of the 40 motifs (Figure 3.A). We observed that the cell types/tissues with the same origin tend to be clustered together, such as embryo muscular cells, immune cells and cells from the digestive tract.

Obviously, some motifs are universally important in the 154 cells/tissues while some are more specific. We calculated the normalized standard deviation (NSD, standard deviation divided by absolute mean value) of linear model weights for each motif. We classified NSD into 3 levels of low (<0.5), medium (0.5, 1) and high (>1) (Figure 3.B). Among the 20 positive weighted motifs, 60% had a low NSD, 30% a medium NSD and 10% a high NSD. Among the 20 negative weighted motifs, 65% had a low NSD, 15% a medium NSD and 20% a high NSD. For both negative and positive weight motifs, most of them (25 out of 40) had a low NSD, suggesting their broad importance in regulating gene expression across diverse cell/tissue types. Interestingly, 13 out of the 15 medium and high NSD motifs had top GO-terms in regulation and stimulus-response pathways. A possible explanation is that regulation and stimulus-response genes have a high variability of expression level in different cell types and thus are regulated by their associated motifs in very different ways.

**Discussion**

Here we have conducted a proof-of-concept study to demonstrate a new strategy to derive linear formulas to quantify the relationship between biological data. We showed that a linear model of DNA motifs could predict the median gene expression levels with a performance comparable to deep neural networks. Conventional approaches to interpret a neural network are to identify the important features and the contribution of each feature to prediction is evaluated by the reduction of prediction performance if a feature is removed. While these approaches are powerful, they do not quantify the relationship between individual features and the predicted variable(s) with an explicit linear regression model. Our study provides a possible solution by combining neural networks with linear models to create an interpretable and high performance model to quantitatively assess the individual DNA motifs' regulation on gene expression.

Linear relationship is the simplest formula to define biology laws and likely easiest to achieve during evolution of the biological systems. One key step of deriving the linear formula is to find appropriate combinations of the input features, which has not been well explored by deep learning approaches. A conventional neural network model using DNA motifs to predict gene expression only achieved a Pearson R=0.515. In comparison, the contextual regression model uncovered features capable of predicting outputs in a linear model with a Pearson R=0.654, comparable to pure deep learning models directly trained from the promoter sequences (Pearson R=0.668). This observation shows the unique advantage of contextual regression on simultaneously optimizing prediction accuracy and assessing the feature contribution, which enforces direct learning of the input features (motifs) fit to a linear formula for predicting outputs (gene expressions).

The strategy we developed here can be readily applied to deriving formulas to quantify the relationship between DNA sequences and different biological functions such as histone modifications, TF binding and open chromatin. This study exemplifies a possible next step of building interpretable models leading to derivation of analytical formulas to decipher the information encoded in the human genome and explain biology that has not been achieved by conventional deep learning models. Additional possible extension of this approach includes conducting symbolic regression on the identified features (motifs) from the CR model to derive non-linear formulas.

Using this approach, we have uncovered 300 motifs that are predictive of the median gene expressions with a linear model achieving similar accuracy as a deep neural network model. The weights of the motifs in the linear model naturally quantify their importance in regulating gene expression. Such a parts list of regulatory motifs in the promoters include a majority of new motifs that are more predictive of gene expression than the known motifs, suggesting unknown regulatory mechanisms implemented in the promoters. Furthermore, we showed these 300 motifs are predictive of cell-type specific gene expression particularly in the embryo cell types/tissues while performing less well on adult cell types/tissues, indicating the increasing importance of enhancers and epigenetic modification during development on controlling gene expressions.

**Acknowledgement**

This work was partially supported by NIH (R01HG009626 to W.W.).

**Supplementary Methods**

**Data**
Data for model training contained 18,377 genes and was directly downloaded from the Agarwal and Shendure study[5] (https://krishna.gs.washington.edu/content/members/vagar/Xpresso/data/datasets/). In this dataset, the genomic sequences of size 10kbp up and downstream of TSSs were converted to one-hot coded sequences. The expression of each gene was downloaded from the Epigenomics Roadmap Consortium[46]. RPKM was computed from RNA-seq and transformed by $x\_ = \log_{10}(x+0.1)$ to reduce the right skewness of the data. The expression levels for prediction was the median of mRNA expression levels across 56 cell types because mRNA expression levels are highly correlated (average correlation of 0.78) between different cell types[5].

**Contextual Regression Model Configuration**
The model was designed to find 8-mer motifs and a total of 64 motifs (filters) were allowed. The model is composed of a two-branched neural network. The sequence is first input into a convolutional layer with 4 input (representing the possible base-pairs) and 64 output channels (representing 64 motifs). Then the output is fed into 2 branches: the first branch applies a sigmoid function to binarize output to either 1 (which represents motif found) or 0 (motif not found), and then sum across the sequence to generate the count of each motif. The other branch has the ordinary structure of a deep convolutional neural network to convert the input into contextual weight of size 64, which corresponds to the local linear weight of each motif found by the other branch. Then the outputs are dot producted with motif counts output from the other branch to generate predictions of expression level. The detailed structure of the model is in Figure 1SA. The model was trained with a stochastic gradient descent optimizer with batch size of 128, learning rate of 0.0005 and momentum of 0.9 for a max epoch number of 100 with early stopping. The model training was carried out using a google colab machine with tensorflow version 2.8 and keras version 2.8. The code can be found at https://github.com/Wang-lab-UCSD/Motif_Finding_Contextual_Regression

**Linear Model Simplification**
We used the LinearRegression function from sklearn package version 0.24.2. The model was trained with least squared loss which comes from the analytical solution of the linear model.
For the global linear model of contextual regression, the output from the motif_counts layer is used as the input. For the global linear model of Xpresso, the output from the first convolutional layer activated by the ReLU activation function and then summed over each (-200bp, +200bp) region (activation counts) is used as the input.

**Individual Cell Line Computing**
The 154 individual cell/tissue type data were the RPKM data downloaded from The Epigenomic Roadmap Project (https://egg2.wustl.edu/) and ENCODE (https://www.encodeproject.org/). We applied the same transformation $x\_ = \log_{10}(x+0.1)$ to reduce the right skewness of the data which is the same preprocessing as in the ref.[5].

**Individual Cell Line Clustering**
The Individual cell line clustering was done using pheatmap package in R.

**Motif Scanning with FIMO**

Default parameters were used for the FIMO search. The positional weight matrices (PWMs) used for FIMO search come from JASPAR 2022 Core Vertebrate set.

**Top Motif Selection**

The motif selection was done using the recursive feature elimination from the sklearn package with DecisionTreeRegressor from the same package as estimator.

**GO-term Analysis**

Go-term analysis was done using http://geneontology.org/ (2022-07-01)

**Statistical Test**

The statistical test on Pearson R of individual cell line models were performed using the stats module from the scipy package

**Figures:**

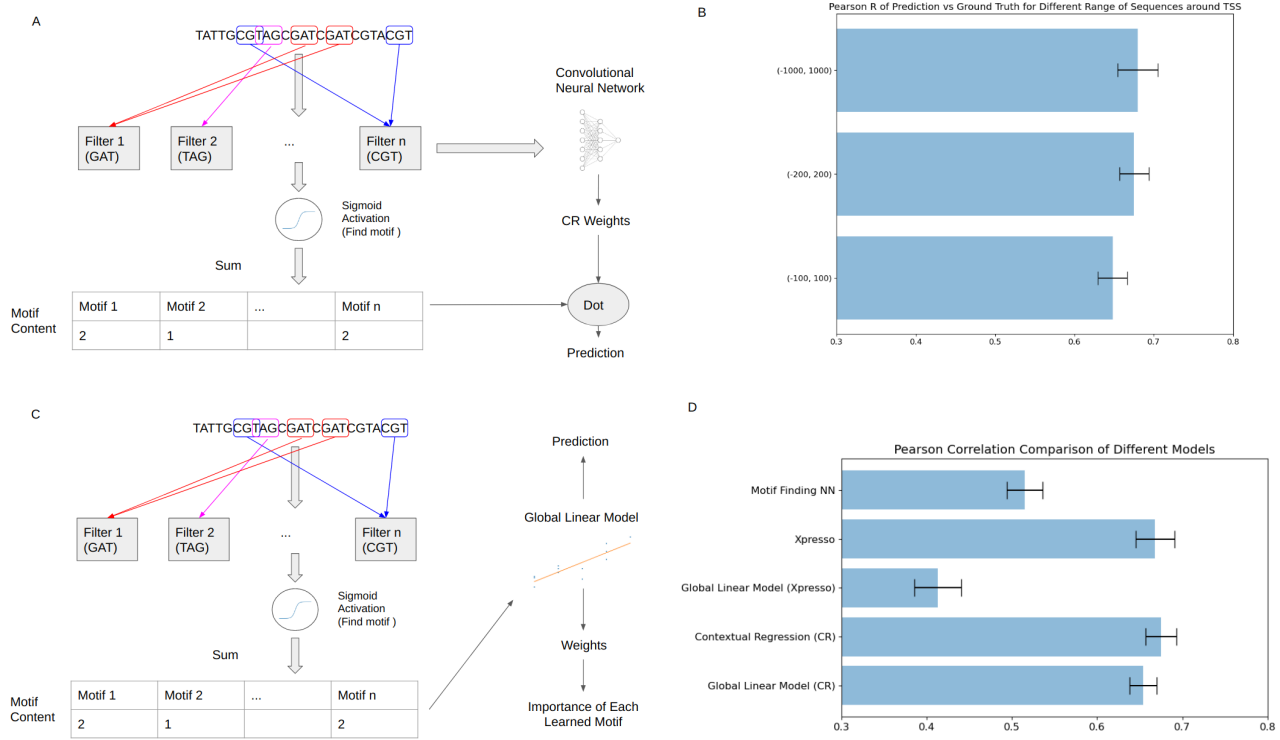

Figure 1. A. A new contextual regression model to predict expressions from promoter sequences. B. Accuracy of the contextual regression model using sequences of different sizes around TSS. C. Building a global linear model from a contextual regression output. D. Accuracy of different models on using (-200, 200) sequence around TSS.

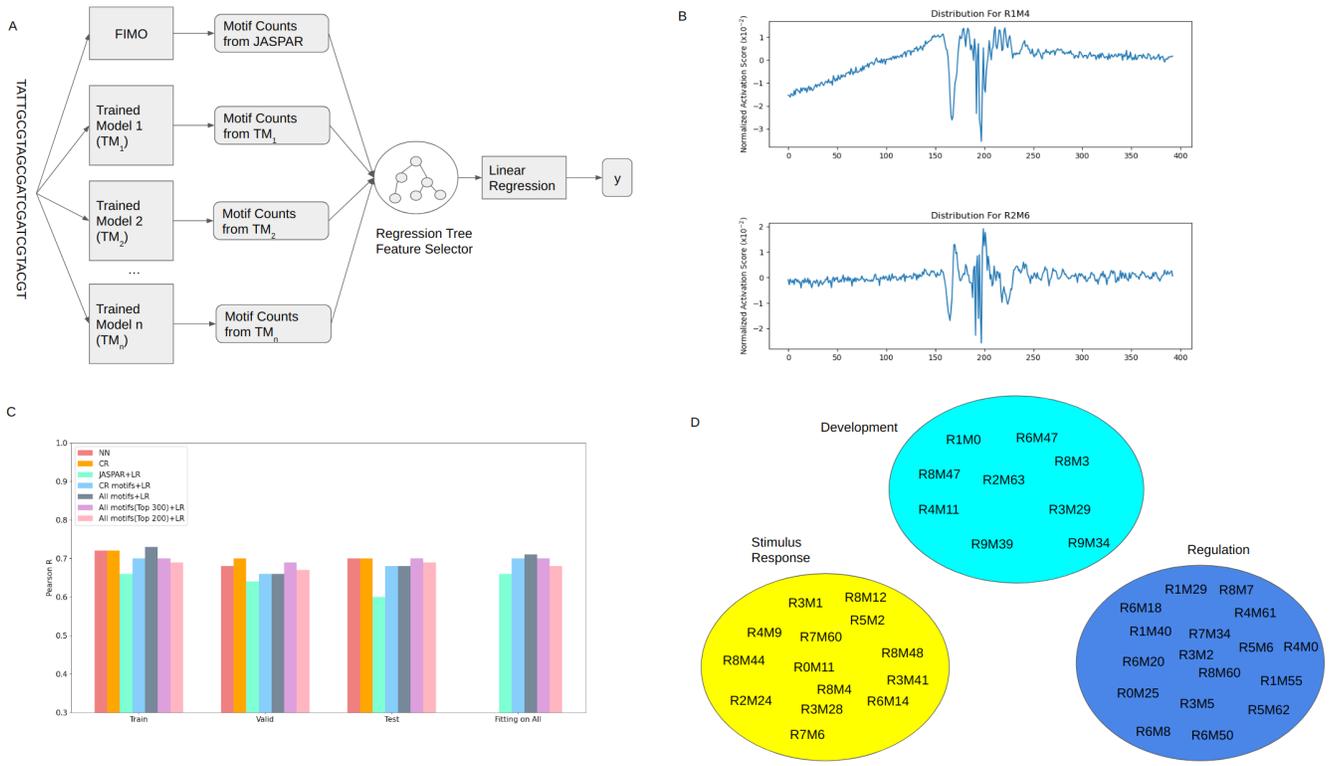

Figure 2. A. Selection of 300 promoter motifs important for predicting gene expression. B. Examples of the two typical distributions of motifs around TSS (R1M4 and R2M6) C. Accuracy comparison of different models on train, valid, test and all set D. The top 20 positive and negative weighted motifs grouped by their associated GO term categories.

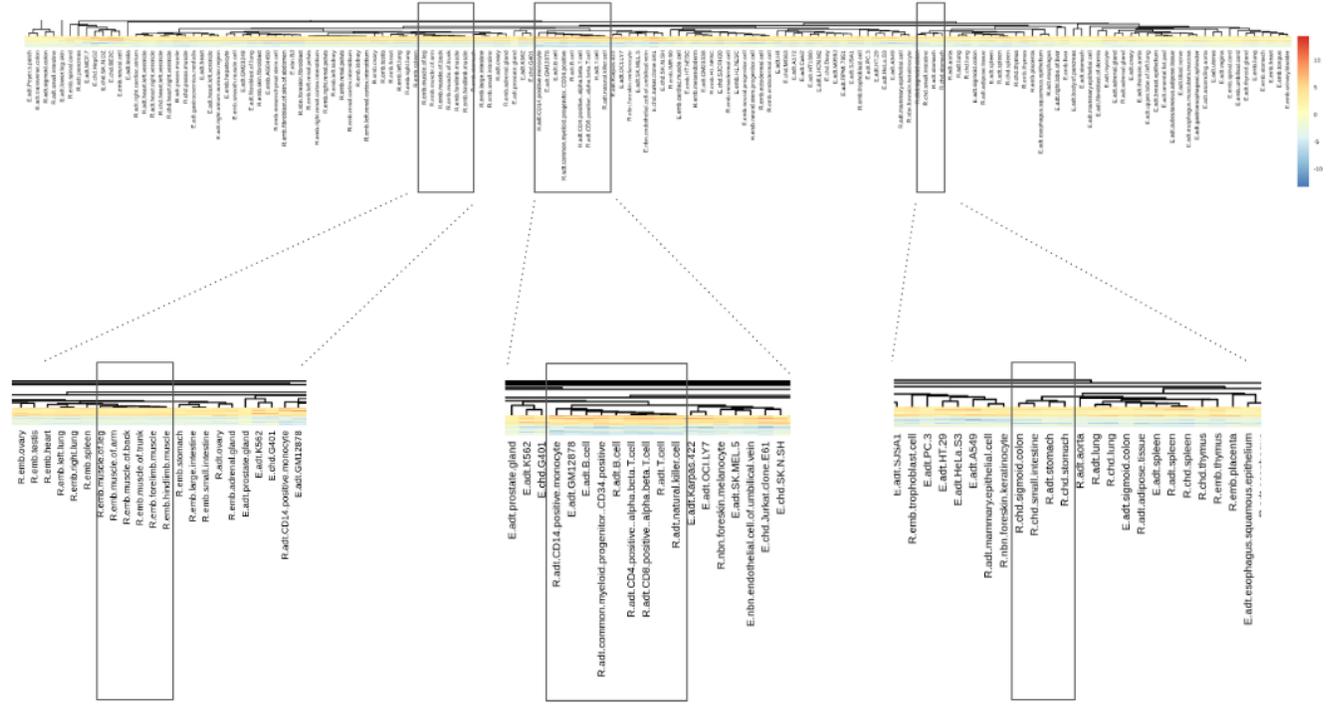

| | Mean Pearson R | Number of Samples | t-test p value vs adult |
|---|---|---|---|
| Adult | 0.593 | 77 | N/A |
| Child | 0.620 | 18 | 0.02 |
| New Born | 0.648 | 4 | N/A |
| Embryo | 0.636 | 54 | 5.6e-7 |

C

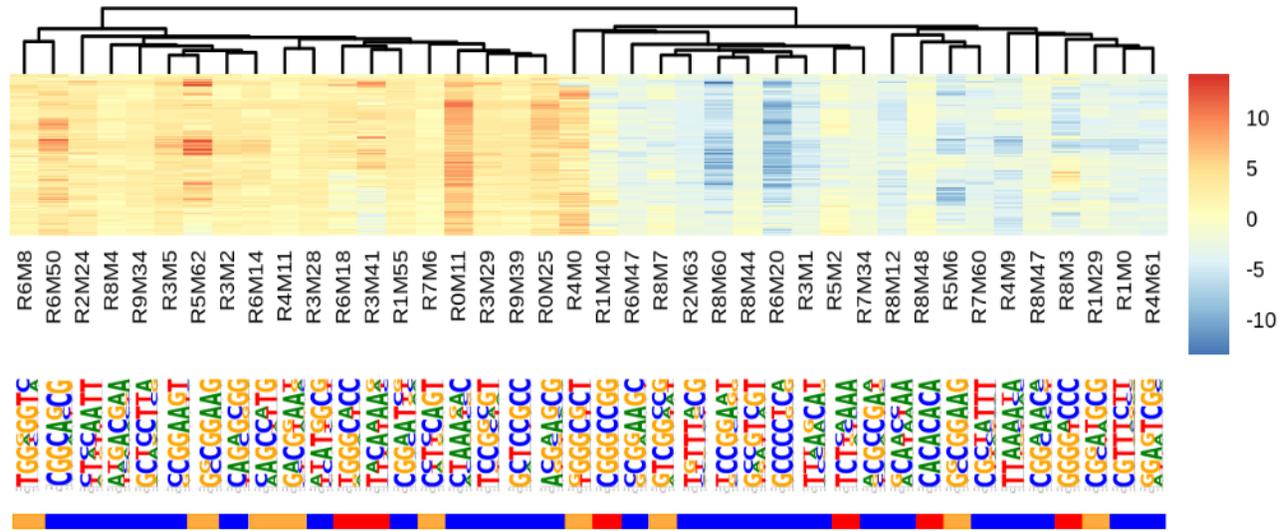

Figure 3. A. Hierarchical clustering of cell/tissue types by the linear model weights of the selected 300 motifs. Noticeable clusters of similar cell/tissue types are shown in amplification. B. Mean value of Pearson correlation, number of samples for the four different developmental stages and their one-sided t test p value vs the adult cells/tissues. C. The linear model weights of the 40 motifs in different cell/tissue types. The NSD level of each motif is marked below the motif sequence: blue represents low NSD, orange represents medium NSD and red represents high NSD.

**Supplementary Figures:**

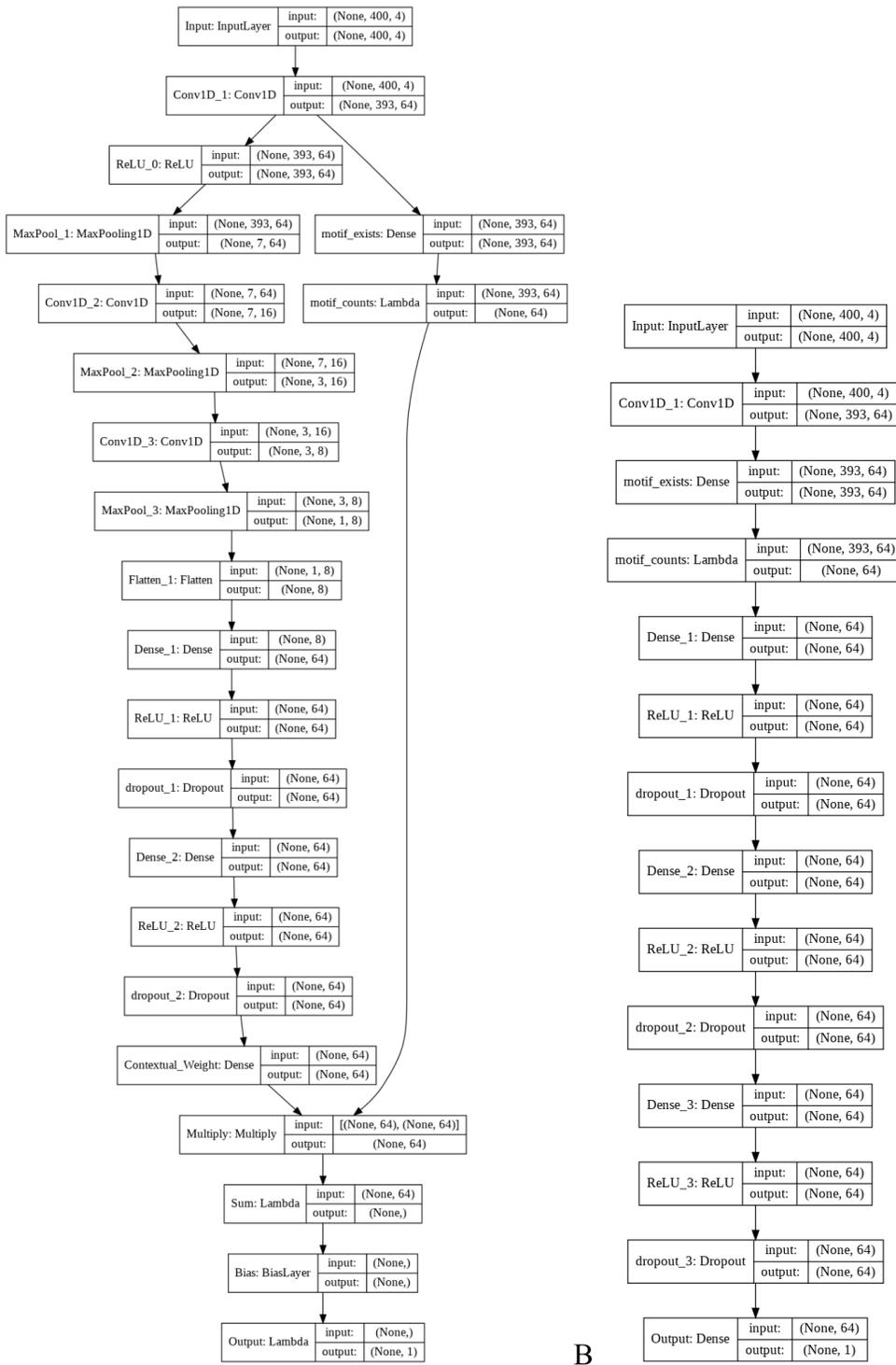

**Figure 1S**. A. Structure of the contextual regression model, the left branch is the contextual weight branch and the right branch is the motif finding branch. B. Structure of the motif finding NN

**Table 1S.** Top 20 negative and positive weighted motifs, with their PWM plot, distribution (average by basepair) around TSS and the top 5 associated GO terms.

| Negative Weight Motif Symbol | Motif PWM | Distribution | Biological Process (Top 5 FDR) | Molecular Function (Top 5 FDR) |
|---|---|---|---|---|
| R1M0 | 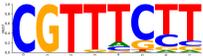 | 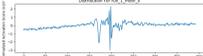 | nervous system development (GO:0007399), negative regulation of cellular process (GO:0048523), regulation of cell junction assembly (GO:1901888), multicellular organism development (GO:0007275), carboxylic acid metabolic process (GO:0019752) | N/A |
| R1M29 | 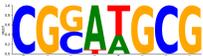 | 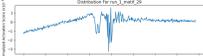 | regulation of signaling (GO:0023051), regulation of cell communication (GO:0010646), regulation of cellular process (GO:0050794), signaling (GO:0023052), regulation of signal transduction (GO:0009966) | nucleoside-triphosphatase regulator activity (GO:0060589), GTPase regulator activity (GO:0030695), protein serine/threonine/tyrosine kinase activity (GO:0004712), protein serine/threonine/tyrosine kinase activity (GO:0004712), guanyl-nucleotide exchange factor activity (GO:0005085) |
| R1M40 | 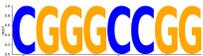 | 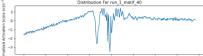 | biological regulation | binding (GO:0005488), |

| | | | (GO:0065007), regulation of cellular process (GO:0050794), negative regulation of cellular process (GO:0048523), regulation of molecular function (GO:0065009), regulation of signaling (GO:0023051) | protein binding (GO:0005515), nucleoside-triphosphatase regulator activity (GO:0060589), GTPase regulator activity (GO:0030695), molecular_function (GO:0003674) |
|---|---|---|---|---|
| R2M63 | 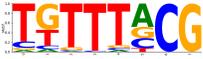 | 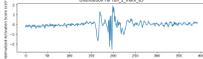 | nervous system development (GO:0007399), synapse organization (GO:0050808), system development (GO:0048731), multicellular organismal process (GO:0032501), multicellular organism development (GO:0007275) | N/A |
| R3M1 | 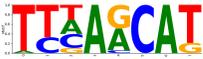 | 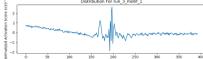 | detection of chemical stimulus (GO:0009593), detection of chemical stimulus involved in sensory perception (GO:0050907), detection of chemical stimulus involved in sensory perception of smell (GO:0050911), sensory perception of smell | olfactory receptor activity (GO:0004984), G protein-coupled receptor activity (GO:0004930), transmembrane signaling receptor activity (GO:0004888), signaling receptor activity (GO:0038023), molecular transducer activity (GO:0060089) |

| | | | (GO:0007608), sensory perception of chemical stimulus (GO:0007606) | |
| --- | --- | --- | --- | --- |
| R4M9 | 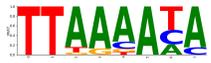 | 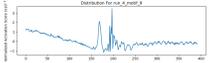 | detection of chemical stimulus (GO:0009593), detection of chemical stimulus involved in sensory perception (GO:0050907), G protein-coupled receptor signaling pathway (GO:0007186), sensory perception of chemical stimulus (GO:0007606), detection of chemical stimulus involved in sensory perception of smell (GO:0050911) | G protein-coupled receptor activity (GO:0004930), olfactory receptor activity (GO:0004984), transmembrane signaling receptor activity (GO:0004888), signaling receptor activity (GO:0038023), molecular transducer activity (GO:0060089) |
| R4M61 | 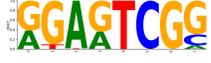 | 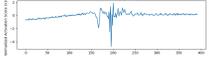 | regulation of transcription DNA-templated (GO:0006355), regulation of developmental process (GO:0050793), regulation of cell communication (GO:0010646), biological regulation (GO:0065007), regulation of Wnt signaling pathway (GO:0030111) | binding (GO:0005488), molecular_function (GO:0003674), nucleoside-triphosphatase regulator activity (GO:0060589), GTPase regulator activity (GO:0030695), enzyme activator activity (GO:0008047) |
| R5M2 | 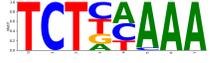 | 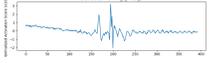 | detection of chemical stimulus | olfactory receptor activity |

| | | | involved in sensory perception (GO:0050907), detection of chemical stimulus (GO:0009593), detection of chemical stimulus involved in sensory perception of smell (GO:0050911), sensory perception of smell (GO:0007608), sensory perception of chemical stimulus (GO:0007606) | (GO:0004984), G protein-coupled receptor activity (GO:0004930), transmembrane signaling receptor activity (GO:0004888), signaling receptor activity (GO:0038023), molecular transducer activity (GO:0060089) |
|---|---|---|---|---|
| R5M6 | 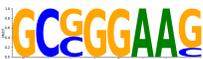 | 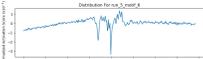 | regulation of cell communication (GO:0010646), regulation of biological process (GO:0050789), regulation of cellular process (GO:0050794), regulation of signaling (GO:0023051), biological regulation (GO:0065007) | transcription coregulator activity (GO:0003712), guanyl-nucleotide exchange factor activity (GO:0005085), nucleoside-triphosphatase regulator activity (GO:0060589), GTPase regulator activity (GO:0030695), protein binding (GO:0005515) |
| R6M20 | 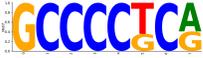 | 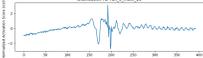 | regulation of signaling (GO:0023051), regulation of cell communication (GO:0010646), regulation of signal transduction (GO:0009966), regulation of response to | protein serine/threonine/tyrosine kinase activity (GO:0004712), protein kinase activity (GO:0004672), phosphotransferase activity alcohol group as acceptor |

| | | | stimulus (GO:0048583), phosphorylation (GO:0016310) | (GO:0016773), kinase activity (GO:0016301), transferase activity transferring phosphorus-containing groups (GO:0016772) |
|---|---|---|---|---|
| R6M47 | 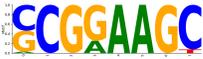 | 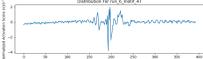 | system development (GO:0048731), developmental process (GO:0032502), multicellular organism development (GO:0007275), anatomical structure development (GO:0048856), regulation of biological process (GO:0050789) | protein tyrosine kinase binding (GO:1990782), protein binding (GO:0005515), binding (GO:0005488), olfactory receptor activity (GO:0004984), sequence-specific DNA binding (GO:0043565) |
| R7M34 | 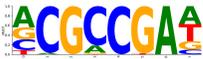 | 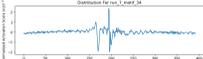 | regulation of signaling (GO:0023051), regulation of cell communication (GO:0010646), regulation of signal transduction (GO:0009966), regulation of response to stimulus (GO:0048583), multicellular organism development (GO:0007275) | protein binding (GO:0005515), signaling receptor binding (GO:0005102), binding (GO:0005488), olfactory receptor activity (GO:0004984), signaling receptor regulator activity (GO:0030545) |
| R7M60 | 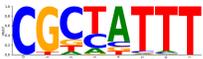 | 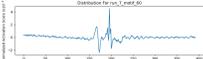 | detection of chemical stimulus involved in sensory perception | olfactory receptor activity (GO:0004984), G protein-coupled |

| | | | (GO:0050907), detection of chemical stimulus involved in sensory perception of smell (GO:0050911), sensory perception of smell (GO:0007608), detection of chemical stimulus (GO:0009593), sensory perception of chemical stimulus (GO:0007606) | receptor activity (GO:0004930), transmembrane signaling receptor activity (GO:0004888), signaling receptor activity (GO:0038023), molecular transducer activity (GO:0060089) |
|---|---|---|---|---|
| R8M3 | 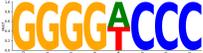 | 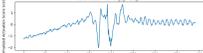 | system development (GO:0048731), regulation of molecular function (GO:0065009), protein modification process (GO:0036211), biological regulation (GO:0065007), cellular protein modification process (GO:0006464) | catalytic activity acting on a protein (GO:0140096), binding (GO:0005488), molecular_function (GO:0003674), protein binding (GO:0005515), ion binding (GO:0043167) |
| R8M7 | 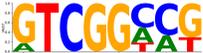 | 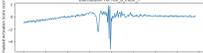 | signaling (GO:0023052), regulation of signaling (GO:0023051), regulation of cell communication (GO:0010646), regulation of molecular function (GO:0065009), system development | nucleoside-triphosphatase regulator activity (GO:0060589), GTPase regulator activity (GO:0030695), protein kinase activity (GO:0004672), GTPase activator activity (GO:0005096), |
| | | | (GO:0048731) | protein serine/threonine/tyrosine kinase activity (GO:0004712) |
| R8M12 | 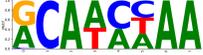 | 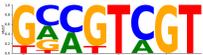 | G protein-coupled receptor signaling pathway (GO:0007186), detection of chemical stimulus involved in sensory perception (GO:0050907), detection of chemical stimulus (GO:0009593), sensory perception of chemical stimulus (GO:0007606), detection of stimulus involved in sensory perception (GO:0050906) | G protein-coupled receptor activity (GO:0004930), olfactory receptor activity (GO:0004984), transmembrane signaling receptor activity (GO:0004888), signaling receptor activity (GO:0038023), molecular transducer activity (GO:0060089) |
| R8M44 | 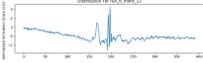 | 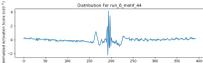 | sensory perception of chemical stimulus (GO:0007606), detection of chemical stimulus involved in sensory perception of smell (GO:0050911), detection of chemical stimulus involved in sensory perception (GO:0050907), detection of chemical stimulus (GO:0009593), sensory perception of smell (GO:0007608) | olfactory receptor activity (GO:0004984), G protein-coupled receptor activity (GO:0004930), signaling receptor activity (GO:0038023), molecular transducer activity (GO:0060089), transmembrane signaling receptor activity (GO:0004888) |

| R8M47 | 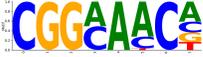 | 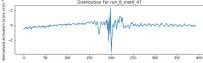 | multicellular organism development (GO:0007275), regulation of cell communication (GO:0010646), regulation of cellular process (GO:0050794), regulation of developmental process (GO:0050793), regulation of signaling (GO:0023051) | transcription regulator activity (GO:0140110), protein binding (GO:0005515), adenyl nucleotide binding (GO:0030554), binding (GO:0005488), olfactory receptor activity (GO:0004984) |
|---|---|---|---|---|
| R8M48 | 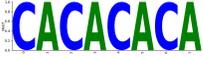 | 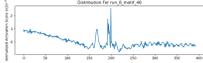 | detection of chemical stimulus involved in sensory perception (GO:0050907), detection of chemical stimulus (GO:0009593), detection of chemical stimulus involved in sensory perception of smell (GO:0050911), sensory perception of smell (GO:0007608), sensory perception of chemical stimulus (GO:0007606) | olfactory receptor activity (GO:0004984), G protein-coupled receptor activity (GO:0004930), transmembrane signaling receptor activity (GO:0004888), signaling receptor activity (GO:0038023), molecular transducer activity (GO:0060089) |
| R8M60 | 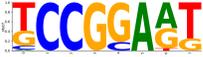 | 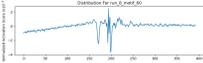 | negative regulation of cellular process (GO:0048523), regulation of molecular function (GO:0065009), regulation of cellular process | protein binding (GO:0005515), binding (GO:0005488), molecular_function (GO:0003674), regulation of signaling (GO:0023051), |

| | | | (GO:0050794), negative regulation of biological process (GO:0048519), regulation of signal transduction (GO:0009966) | protein modification process (GO:0036211) |

| Positive Weight Motif Symbol | Motif PWM | Distribution | Biological Process | Molecular Function |
|---|---|---|---|---|
| R0M11 | 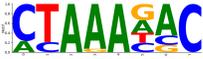 | 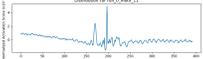 | G protein-coupled receptor signaling pathway (GO:0007186), detection of chemical stimulus involved in sensory perception (GO:0050907), detection of chemical stimulus (GO:0009593), sensory perception of chemical stimulus (GO:0007606), detection of stimulus involved in sensory perception (GO:0050906) | G protein-coupled receptor activity (GO:0004930), transmembrane signaling receptor activity (GO:0004888), olfactory receptor activity (GO:0004984), signaling receptor activity (GO:0038023), molecular transducer activity (GO:0060089) |
| R0M25 | 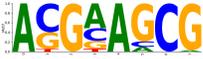 | 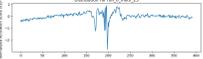 | regulation of cellular process (GO:0050794), regulation of developmental process (GO:0050793), regulation of biological process | transcription regulator activity (GO:0140110), olfactory receptor activity (GO:0004984), nucleoside-triphosphatase regulator activity |

| | | | (GO:0050789), regulation of signaling (GO:0023051), regulation of cell communication (GO:0010646) | (GO:0060589), GTPase regulator activity (GO:0030695), guanyl-nucleotide exchange factor activity (GO:0005085) |
|---|---|---|---|---|
| R1M55 | 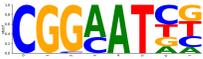 | 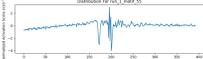 | regulation of signaling (GO:0023051), regulation of cell communication (GO:0010646), regulation of cellular process (GO:0050794), regulation of signal transduction (GO:0009966), regulation of biological process (GO:0050789) | nucleoside-triphosphatase regulator activity (GO:0060589), GTPase regulator activity (GO:0030695), ion binding (GO:0043167), protein serine/threonine/tyrosine kinase activity (GO:0004712), guanyl-nucleotide exchange factor activity (GO:0005085) |
| R2M24 | 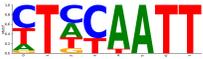 | 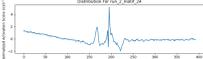 | detection of chemical stimulus (GO:0009593), detection of chemical stimulus involved in sensory perception (GO:0050907), detection of chemical stimulus involved in sensory perception of smell (GO:0050911), sensory perception of smell (GO:0007608), sensory perception of chemical stimulus (GO:0007606) | olfactory receptor activity (GO:0004984), G protein-coupled receptor activity (GO:0004930), transmembrane signaling receptor activity (GO:0004888), signaling receptor activity (GO:0038023), molecular transducer activity (GO:0060089) |
| R3M2 | 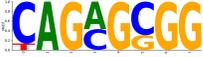 | 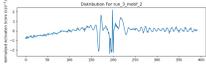 | regulation of developmental process (GO:0050793), regulation of cell communication (GO:0010646), regulation of signaling (GO:0023051), system development (GO:0048731), regulation of signal transduction (GO:0009966) | protein binding (GO:0005515), binding (GO:0005488), molecular_function (GO:0003674), transcription coregulator activity (GO:0003712), protein domain specific binding (GO:0019904) |
| R3M5 | 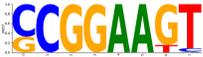 | 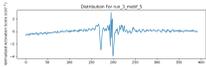 | biological regulation (GO:0065007), regulation of signaling (GO:0023051), regulation of cell communication (GO:0010646), regulation of cellular process (GO:0050794), regulation of biological process (GO:0050789) | binding (GO:0005488), nucleoside-triphosphatase regulator activity (GO:0060589), protein serine/threonine/tyrosine kinase activity (GO:0004712), phosphotransferase activity alcohol group as acceptor (GO:0016773), protein binding (GO:0005515) |
| R3M28 | 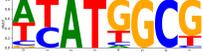 | 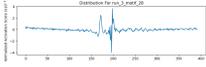 | detection of chemical stimulus (GO:0009593), detection of chemical stimulus involved in sensory perception (GO:0050907), sensory perception of chemical stimulus (GO:0007606), detection of chemical stimulus | olfactory receptor activity (GO:0004984), G protein-coupled receptor activity (GO:0004930), transmembrane signaling receptor activity (GO:0004888), signaling receptor activity (GO:0038023), molecular |
| | | | involved in sensory perception of smell (GO:0050911), detection of stimulus involved in sensory perception (GO:0050906) | transducer activity (GO:0060089) |
| R3M29 | 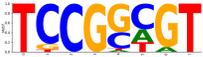 | 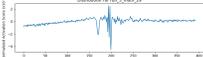 | tube development (GO:0035295), multicellular organism development (GO:0007275), system development (GO:0048731), tube morphogenesis (GO:0035239), regulation of cellular process (GO:0050794) | catalytic activity acting on a protein (GO:0140096), growth factor binding (GO:0019838), binding (GO:0005488), metal ion binding (GO:0046872), cation binding (GO:0043169) |
| R3M41 | 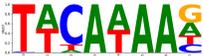 | 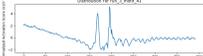 | detection of chemical stimulus (GO:0009593), detection of chemical stimulus involved in sensory perception (GO:0050907), sensory perception of chemical stimulus (GO:0007606), detection of stimulus (GO:0051606), detection of stimulus involved in sensory perception (GO:0050906) | G protein-coupled receptor activity (GO:0004930), olfactory receptor activity (GO:0004984), transmembrane signaling receptor activity (GO:0004888), signaling receptor activity (GO:0038023), molecular transducer activity (GO:0060089) |
| R4M0 | 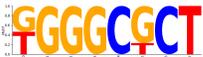 | 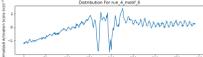 | biological regulation | binding (GO:0005488), |

| | | | (GO:0065007), regulation of cellular process (GO:0050794), biological_process (GO:0008150), regulation of molecular function (GO:0065009), regulation of biological process (GO:0050789) | protein binding (GO:0005515), molecular_function (GO:0003674), protein kinase activity (GO:0004672), modification-dependent protein binding (GO:0140030) |
|---|---|---|---|---|
| R4M11 | 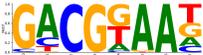 | 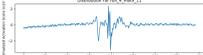 | system development (GO:0048731), multicellular organism development (GO:0007275), developmental process (GO:0032502), anatomical structure development (GO:0048856), regulation of cellular process (GO:0050794) | enzyme regulator activity (GO:0030234), protein binding (GO:0005515), GTPase activator activity (GO:0005096), nucleoside-triphosphatase regulator activity (GO:0060589), olfactory receptor activity (GO:0004984) |
| R5M62 | 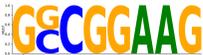 | 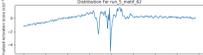 | regulation of molecular function (GO:0065009), system development (GO:0048731), regulation of cellular process (GO:0050794), biological regulation (GO:0065007), regulation of developmental process (GO:0050793) | protein binding (GO:0005515), binding (GO:0005488), molecular_function (GO:0003674), nucleoside-triphosphatase regulator activity (GO:0060589), protein phosphorylated amino acid binding (GO:0045309) |

| R6M8 | 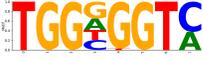 | 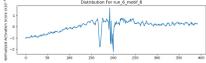 | biological regulation (GO:0065007), biological_process (GO:0008150), regulation of signaling (GO:0023051), regulation of cell communication (GO:0010646), regulation of cellular process (GO:0050794) | protein binding (GO:0005515), binding (GO:0005488), molecular_function (GO:0003674), metal ion binding (GO:0046872), catalytic activity acting on a protein (GO:0140096) |
|---|---|---|---|---|
| R6M14 | 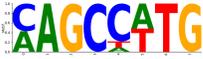 | 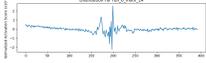 | G protein-coupled receptor signaling pathway (GO:0007186), detection of chemical stimulus involved in sensory perception (GO:0050907), detection of chemical stimulus (GO:0009593), sensory perception of chemical stimulus (GO:0007606), detection of stimulus involved in sensory perception (GO:0050906) | G protein-coupled receptor activity (GO:0004930), olfactory receptor activity (GO:0004984), transmembrane signaling receptor activity (GO:0004888), signaling receptor activity (GO:0038023), molecular transducer activity (GO:0060089) |
| R6M18 | 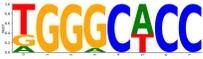 | 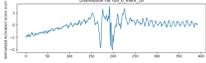 | regulation of molecular function (GO:0065009), system development (GO:0048731), nervous system development (GO:0007399), biological regulation (GO:0065007), | nucleoside-triphosphatase regulator activity (GO:0060589), protein binding (GO:0005515), protein kinase activity (GO:0004672), binding (GO:0005488), molecular_functio |

| | | | negative regulation of cellular process (GO:0048523) | n (GO:0003674) |
|---|---|---|---|---|
| R6M50 | 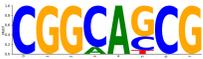 | 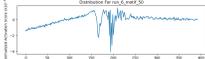 | regulation of signaling (GO:0023051), regulation of cell communication (GO:0010646), regulation of cellular process (GO:0050794), signaling (GO:0023052), regulation of signal transduction (GO:0009966) | protein serine/threonine/tyrosine kinase activity (GO:0004712), protein serine kinase activity (GO:0106310), protein kinase activity (GO:0004672), phosphotransferase activity alcohol group as acceptor (GO:0016773), molecular_function (GO:0003674) |
| R7M6 | 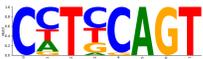 | 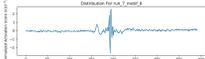 | sensory perception of chemical stimulus (GO:0007606), detection of chemical stimulus involved in sensory perception of smell (GO:0050911), detection of chemical stimulus involved in sensory perception (GO:0050907), detection of chemical stimulus (GO:0009593), sensory perception of smell (GO:0007608) | olfactory receptor activity (GO:0004984), signaling receptor activity (GO:0038023), molecular transducer activity (GO:0060089), transmembrane signaling receptor activity (GO:0004888), G protein-coupled receptor activity (GO:0004930) |
| R8M4 | 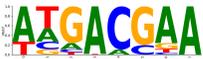 | 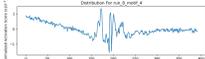 | detection of chemical stimulus involved in sensory perception | G protein-coupled receptor activity (GO:0004930), olfactory receptor |

| | | | | activity (GO:0004984), transmembrane signaling receptor activity (GO:0004888), signaling receptor activity (GO:0038023), molecular transducer activity (GO:0060089) |
| --- | --- | --- | --- | --- |
| | | | (GO:0050907), detection of chemical stimulus (GO:0009593), G protein-coupled receptor signaling pathway (GO:0007186), sensory perception of chemical stimulus (GO:0007606), detection of stimulus involved in sensory perception (GO:0050906) | |
| R9M34 | 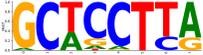 | 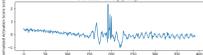 | multicellular organismal process (GO:0032501), nervous system development (GO:0007399), multicellular organism development (GO:0007275), developmental process (GO:0032502), system process (GO:0003008) | olfactory receptor activity (GO:0004984), signaling receptor activity (GO:0038023), transmembrane signaling receptor activity (GO:0004888), molecular transducer activity (GO:0060089), DNA-binding transcription activator activity (GO:0001216) |
| R9M39 | 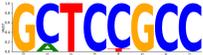 | 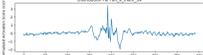 | multicellular organism development (GO:0007275), system development (GO:0048731), nervous system development (GO:0007399), anatomical structure development | binding (GO:0005488), kinase activity (GO:0016301), transferase activity transferring phosphorus-containing groups (GO:0016772), protein kinase activity (GO:0004672), phosphotransferas |

| | | | (GO:0048856), regulation of signaling (GO:0023051) | e activity alcohol group as acceptor (GO:0016773) |